\documentclass[useAMS,usenatbib]{mnras}
\usepackage{amssymb,amsmath,graphicx}

\def\gsim{\\raise 3pt\hbox{$\rangle$}\kern -8.5pt\raise -2pt\hbox{$\sim$}\ }
\def\lsim{\\raise 3pt\hbox{$\langle$}\kern -8.5pt\raise -2pt\hbox{$\sim$}\ }

\newcommand{\sk}{$\widehat{SK}$~}
\newcommand{\esk}{\widehat{SK}}
\newcommand{\bsk}{$\widehat{SK^*}$~}
\newcommand{\ebsk}{\widehat{SK^*}}

\title[Spectral Kurtosis Statistics of Transient Signals]{Spectral Kurtosis Statistics of Transient Signals}
\author[G. M. Nita ]{G. M. Nita$^{1}$\thanks{E-mail: gnita@njit.edu}\\
$^{1}$Center for Solar-Terrestrial Research, New Jersey
Institute of Technology, Newark, NJ 07102, USA}
\begin{document}
\date{}
\pagerange{\pageref{firstpage}--\pageref{lastpage}} \pubyear{2002}

\maketitle

\label{firstpage}

\begin{abstract}
  We obtain analytical approximations for the expectation and variance of the Spectral Kurtosis estimator in the case of Gaussian and coherent transient time domain signals mixed with a quasi-stationary Gaussian background, which are suitable for practical estimations of their signal-to-noise ratio and duty-cycle relative to the instrumental integration time. We validate these analytical approximations by means of numerical simulations and demonstrate that such estimates are affected by statistical uncertainties that, for a suitable choice of the integration time, may not exceed a few percent. Based on these analytical results, we suggest a multiscale Spectral Kurtosis spectrometer design optimized for real-time detection of transient signals, automatic discrimination based on their statistical signature, and measurement of their properties.
\end{abstract}


\section{Introduction}
The Spectral Kurtosis Estimator (\sk) was originally proposed by \citealp{rfi} as a statistical tool for real-time radio frequency
interference (RFI) detection and excision in a Fast Fourier Transform (FFT) radio spectrograph. The first--ever hardware implementation of an \sk spectrograph,
the Korean Solar Radio Burst Locator \citep[KSRBL,][]{ksrbl}, provided comprehensive experimental data that validated its theoretically expected performance \citep{sks}.

The \sk estimator is a higher order unbiased statistical estimator associated with an accumulated Power Spectral  Density (PSD), which is defined as \citep{sk,gsk},
   \begin{equation}  \label{SK}
   \widehat{SK}=\frac{M+1}{M-1}\Big(\frac{M S_2}{S_1^2}-1\Big)
   \end{equation}
where, at each frequency bin $f_k\;(k=1-N/2)$,
\begin{equation}
\label{sums}
S_1(f_k)=\sum_{i=1}^MP_i(f_k),\;\;S_2(f_k)=\sum_{i=1}^MP_i^2(f_k)
\end{equation}
are sums taken over $M$ raw FFT consecutive PSD estimates and, respectively, their squares.

A remarkable property of the \sk estimator is that, in the case of a pure Gaussian time domain signal, its statistical expectation is unity at each frequency bin, while the power spectrum may have an arbitrary spectral shape. This property gives the \sk estimator the ability to discriminate signals deviating from a Gaussian time domain statistics against arbitrarily shaped astronomical backgrounds, as it is usually the case of the man-made signals producing unwanted RFI contamination of the astronomical signals of interest. Nevertheless, as demonstrated by \citet{rfi}, the \sk estimator may be equally sensitive to narrow band astronomical transient signals such as radio spikes, which might be mistakenly flagged by a blind RFI excision algorithm assuming that all \sk values deviating from unity should be excised from the astronomical signal of interest. However, as demonstrated here, rather than being a limitation of its practical applicability, this particular sensitivity of the \sk estimator to transient signals may be exploited not only to detect them, but also to quantitatively characterize their properties. For this purpose, we analyze the statistical properties of these two special classes of transients, and obtain analytical expressions for the expected mean and variance of their associated \sk estimators as functions of their effective durations and signal-to-noise ratios.
\section{Statistics of Gaussian transients}\label{gauss}
In this section we analyze the statistical properties of  \sk estimator in the case of narrow band Gaussian transient time domain signal mixed with a quasi-stationary Gaussian background. In \S\ref{gaussian_stationary} we provide an overview of the results previously reported by \citet{sk} regarding the expected mean and variance of the \sk estimator associated with a quasi-stationary time domain Gaussian signal. In \S\ref{gauss_transients} we generalize these results and provide an analytical expression for the expectation of the \sk estimator in the case of a transient Gaussian signal mixed with a Gaussian time domain background, which, in the $M\gg1$ limit reduces to the $M\gg1$ limit of an analytical expression previously reported by \citet{rfi}. In addition, we obtain an analytical expression for the variance of the \sk estimator under the same conditions. We validate these theoretical expectations by means of numerical simulations.
\subsection{Quasi-stationary Gaussian signals}\label{gaussian_stationary}
As shown by \citet{sk}, in the case of a quasi-stationary time domain signal obeying a Gaussian statistics, the statistical expectations for the FFT-derived $n^{th}$-powers of the sums $S_1$ and $S_2$ defined by Eqn. \ref{sums} are given by
\begin{eqnarray}
\label{gauss_moments}
E(S_1^n)&=&\frac{(M+n-1)!}{(M-1)!}\mu^n\\\nonumber
E(S_2^n)&=&\frac{ \partial^n}{\partial t^n} \Big[ \sum_{r=0}^n\frac{(2r)!}{r!}t^n\Big]^M\Big|_{t=0}\mu^{2n},
\end{eqnarray}
where $\mu$ represents the mean power of the quasi-stationary Gaussian background at the particular frequency bin considered.

In particular, Eqn. \ref{gauss_moments} provides the expectations
\begin{eqnarray}
\label{gauss_s1mom}
E[S_1(M,\mu)]&=&M\mu\\\nonumber
E[S_1^2(M,\mu)]&=&M(M+1)\mu^2\\\nonumber
E[S_1^3(M,\mu)]&=&M(M+1)(M+2)\mu^3\\\nonumber
E[S_1^4(M,\mu)]&=&M(M+1)(M+2)(M+3)\mu^2
\end{eqnarray}
and
\begin{eqnarray}
\label{gauss_s2mom}
E[S_2(M,\mu)]&=&M\mu^2\\\nonumber
E[S_2^2(M,\mu)]&=&4M(5+M)\mu^4,
\end{eqnarray}
which are needed to compute the expected mean and variance of the \sk estimator.
From Eqn. \ref{SK} immediately follows
\begin{eqnarray}
\label{sk_mean}
E\Big[\widehat{SK}\Big]
=\frac{M+1}{M-1}\Big(M\frac{E[S_2]}{E[S_1^2]}-1\Big),
\end{eqnarray}
and
\begin{eqnarray}
\label{sk2_mean}
E\Big[\widehat{SK}^2\Big]=
\Big(\frac{M+1}{M-1}\Big)^2\Big(M^2\frac{E[S_2^2]}{E[S_1^4]}-2M\frac{E[S_2]}{E[S_1^2]}+1\Big),
\end{eqnarray}
where we made use of the non-trivial identity
\begin{equation}
\label{uncorr}
E\Big[\Big(\frac{S_2}{S_1^2}\Big)^n\Big]=\frac{E[S_2^n]}{E[S_1^{2n}]},
\end{equation}
which holds because, in the case of a normally distributed time domain signal, $S_2/S_1^2$ and $S_1^2$ are uncorrelated random variables \citep{sk}.

Hence, by plugging in the expressions provided by Eqnuations \ref{gauss_s1mom} and \ref{gauss_s2mom},
and writing down $\sigma^2_{\widehat{SK}}=E\Big[\widehat{SK}^2\Big]-E\Big[\widehat{SK}\Big]^2$, immediately follows $E\Big[\widehat{SK}\Big]=1$, and
\begin{eqnarray}
\label{varSK}
&&\sigma^2_{\widehat{SK}}=\frac{4M^2}{(M-1)(M+2)(M+3)}\simeq \frac{4}{M},
\end{eqnarray}
where the approximation is valid for accumulation lengths much larger than unity.
\subsection{Gaussian transients}\label{gauss_transients}
To investigate how the \sk estimator is expected to deviate from unity in the case of a quasi-stationary Gaussian background mixed with a Gaussian time domain signal lasting shorter than the accumulation time, we adopt the model originally considered by \citet{rfi}, in which the mean spectral power of the transient signal is characterized by a signal-to-noise ratio $\rho$ relative to the mean power of the quasi-stationary background, $\mu$, and the transient signal is considered to be effectively present only in an integer fraction $\delta M$ of the $M$ raw PSD estimates contributing to the accumulations $S_1$ and $S_2$. Hence, taking in consideration that $S_1$ and $S_2$ are sums of a series of uncorrelated random variables, a simple binomial expansions leads to
\begin{eqnarray}
\label{S1mom}
E\{S_1^n\}=E\Big\{\Big[S_1[(1-\delta)M,\mu]+S_1[\delta M,(1+\rho)\mu]\Big]^n\Big\}\\\nonumber
=\sum_{k=1}^nC_n^k E\Big[S_1^k[(1-\delta)M,\mu])\Big]E\Big[S_1^{n-k}[\delta M,(1+\rho)\mu]\Big]
\end{eqnarray}
and
\begin{eqnarray}
\label{S2mom}
E\{S_2^n\}=E\Big\{\Big[S_2[(1-\delta)M,\mu]+S_2[\delta M,(1+\rho)\mu]\Big]^n\Big\}\\\nonumber
=\sum_{k=1}^nC_n^k E\Big[S_2^k[(1-\delta)M,\mu]\Big]E\Big[S_2^{n-k}[\delta M,(1+\rho)\mu]\Big],
\end{eqnarray}
which are linear combinations of the expectations given by Equations \ref{gauss_s1mom} and \ref{gauss_s2mom}.

Although the identities given by Eqn. \ref{uncorr} do no longer exactly hold in the case of Gaussian transients, we assume that the expressions given by Eqns. \ref{sk_mean} and \ref{sk2_mean} can still serve as biased estimators of the first and second moments of $\widehat{SK}$. Consequently, after a few algebraical manipulations of Eqn. \ref{sk_mean}, the biased \bsk estimator corresponding to the adopted Gaussian transient model can be written as
\begin{eqnarray}
\label{gauss_sk_full}
\ebsk=1 + \frac{2 (1 - \delta) \delta \rho^2 M^2 }
{(1 + \delta \rho)^2 M^2 + (1 - \delta) \delta \rho^2  M - (1 + 2 \delta \rho + \delta \rho^2)},
\end{eqnarray}
which, for accumulation lengths $M\gg1$,  reduces to
\begin{eqnarray}
\label{gauss_sk_approx}
\ebsk\simeq 1 + \frac{2 (1 - \delta) \delta \rho^2}{(1 + \delta \rho)^2 }.
\end{eqnarray}
As a first self-consistency check of the adopted \bsk approximation, we note that, for any $\rho$, Eqns. \ref{gauss_sk_full} and \ref{gauss_sk_approx} reduce to unity when $\delta=0$ or $\delta=1$, i.e. the cases of no transient signal or, respectively, the mixture of two quasi-stationary Gaussian signals. Therefore, \bsk produces unbiased estimates at both ends of the duty-cycle range. For all other cases, \bsk deviates from unity and reaches a maximum at $\delta=1/(2+\rho)$, i.e.
\begin{equation}
\label{gauss_max_sk}
\ebsk\Big(\frac{1}{2+\rho}\Big)=1+\frac{\rho^2}{2(1+\rho)}.
\end{equation}
Similarly, from Eqn. \ref{sk2_mean}, we obtain an approximation for the variance of \sk that, for $M\gg1$, reduces to
\begin{eqnarray}
\label{gauss_varSK}
\sigma^2_{\ebsk}\simeq\frac{4}{M(1 + \delta \rho)^6 }
   (1+6 \delta \rho + 18 \delta \rho^2 -3 \delta^2 \rho^2 + 20 \delta \rho^3\\\nonumber
   +12 \delta^2 \rho^3 - 12 \delta^3 \rho^3 +5 \delta \rho^4 + 28 \delta^2 \rho^4 \\\nonumber
   -18 \delta^3 \rho^4 + 10 \delta^2 \rho^5 -4 \delta^3 \rho^5 + \delta^3 \rho^6).
\end{eqnarray}
For $\delta=0$ and $\delta=1$, 
Eqn. \ref{gauss_varSK} reduce to the expressions given by Eqn. \ref{varSK}, while in between these extremes, and fixed SNR, $\sigma^2_{\ebsk}$ has a single-peaked duty-cycle dependence.

\begin{figure}
 \centerline{\includegraphics[width=1\columnwidth,clip=]{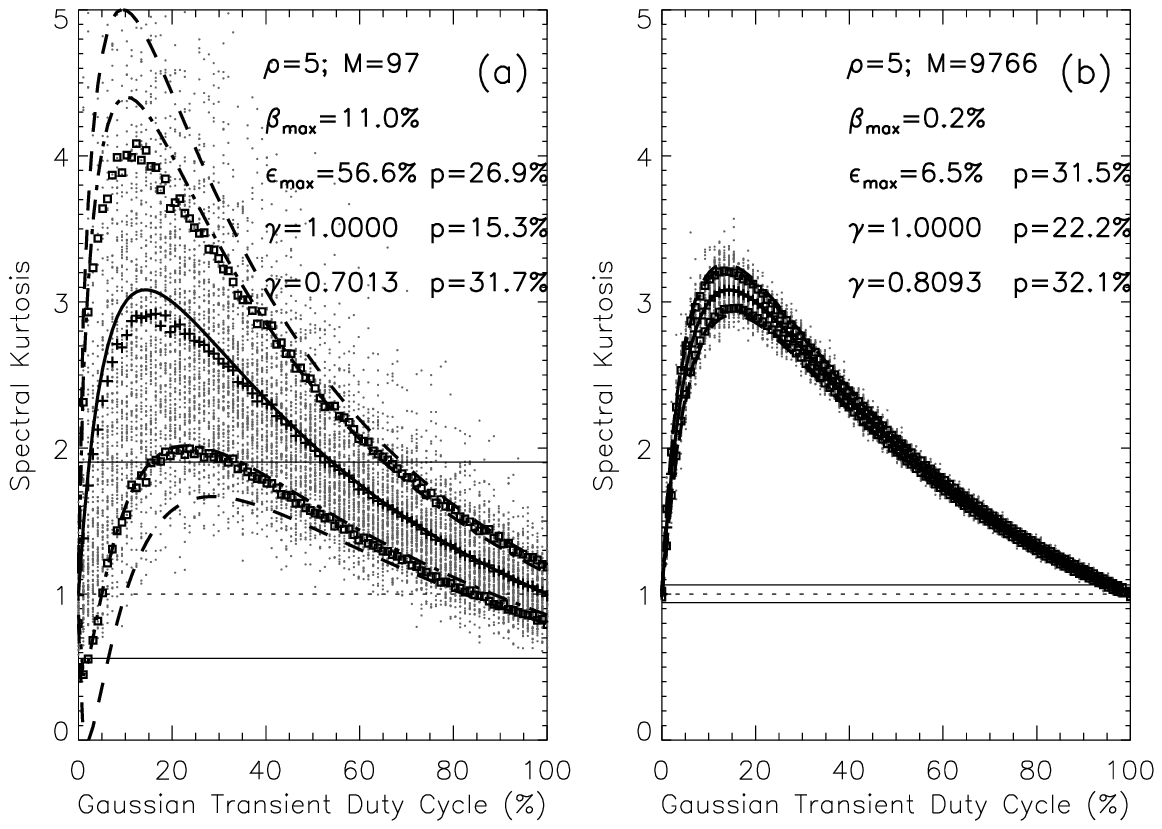}}
 \centerline{\includegraphics[width=1\columnwidth,clip=]{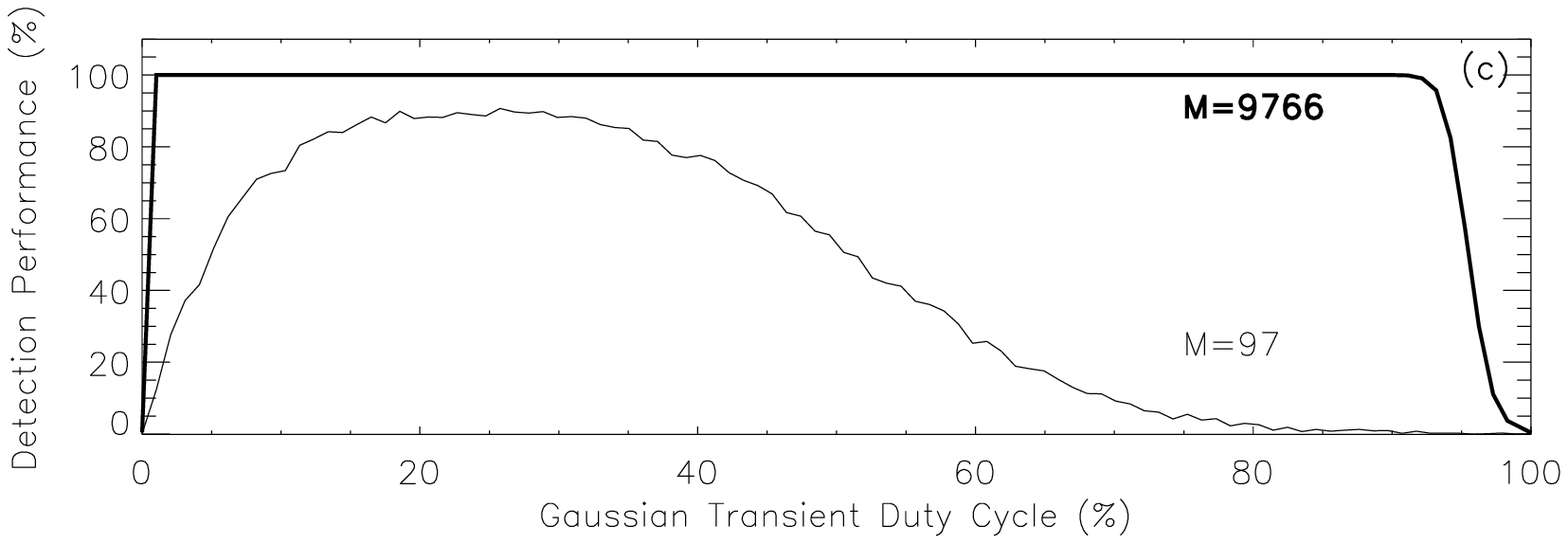}}
 \caption{ \bsk estimator (solid lines) as function of a Gaussian transient duty-cycle for a signal-to-noise ratio $\rho=5$ and two accumulations lengths, $M=97$ (panel a) and $M=9766$ (panel b). The symmetric $\ebsk\pm\sigma_{\ebsk}$ limits provided by Eqn. \ref{gauss_varSK} are overlaid (dashed lines) on top of the corresponding numerically simulated distributions (point symbols). The SDEV--corrected $68.27\%$ probability ranges,  $\ebsk\pm\gamma\sigma_{\ebsk}$, are indicated by the dot-dashed lines. The square symbols indicate the sample standard deviations around the mean of the simulated \sk distribution (plus symbols).  The Pearson Type IV asymmetric detection thresholds, which correspond to standard $0.13499\%$ probabilities of false alarm on each side of the unity \sk expectation (dotted lines) if no transient emission was present, are indicated by horizontal thin lines. The SDEV correction factors $\gamma$, the maximum values reached by the duty-cycle dependent relative standard deviations $\epsilon_{max}$, and the corresponding maximum relative bias of the \sk estimator, $\beta_{max}$, are indicated in each figure inset. Panel (c) displys the percentage of \sk--detected transients as function of their duty-cycle, for $M=97$ (thin line), and $M=9766$ (thick line).} \label{gauss_mc}
 \end{figure}

Fig. \ref{gauss_mc} displays, for a fixed SNR ($\rho\equiv5$), the duty-cycle dependence of the \bsk approximation (solid lines), as well as the $\ebsk\pm\sigma_{\ebsk}$ fluctuation ranges (dashed lines), for two selected accumulation lengths, $M=97$ (panel a) and $M=9766$ (panel b), which correspond to the accumulation lengths of two short--lived prototype instruments equipped with \sk capabilities, the Frequency Agile Solar Radiotelescope Subsystem Testbed \citep[FST,][]{FST} and the Expanded Owens Valley Solar Array Subsystem Testbed \citep[EST,][]{EST}. We compare these theoretical expectations with Monte-Carlo simulations of the $\widehat{SK}(\delta,\rho\equiv5)$ distributions (scattered points) that were obtained by generating $1000$ \sk values for each of the duty-cycle values ranging from $0\%$ to $100\%$ . This comparison demonstrates an overall good agreement of the $\delta$--dependent mean of the simulated \sk distributions, $\langle\esk\rangle$, (cross symbols), with the \bsk predictions (solid lines), which are affected by a $\delta$--dependent bias that reaches its maximum in the vicinity of the \bsk peak, but significantly decreases as the accumulation length $M$ increases.

Fig. \ref{gauss_mc}a,b insets indicate the maximum observed relative bias, $\beta\equiv\ebsk/\langle\esk\rangle-1$, i.e. $\beta_{max}=11.0\%$ for $M=97$, and $\beta_{max}=0.2\%$ for $M=9766$, which we compare with the sample maximum relative standard deviations (SDEV), $ \epsilon\equiv\sqrt{\langle\esk^2\rangle/\langle\esk\rangle^2-1}$, i.e. $\epsilon_{max}=56.6\%$  and $\epsilon_{max}=6.5\%$, respectively. This comparison reveals that not only the maximum \bsk bias is smaller than the absolute relative standard deviation $\epsilon$, but also, for any duty-cycle, it is contained within the sample SDEV range $\langle\esk\rangle(1\pm\epsilon)$ (square symbols). Therefore, we may conclude that the choice of the analytical expression given by Eqn. \ref{gauss_sk_approx} as an biased estimator for the true mean of the \sk distribution may be accurate enough for being employed in practical applications, especially those involving large accumulation lengths M. Nevertheless, despite being consistent with the large scatter of the simulated \sk distribution, the $\ebsk\pm\sigma_{\ebsk}$ ranges (dashed lines) appear to largely overestimate, especially for small values of M, the true variance of the \sk distribution, as estimated by the sample SDEV range indicated by the square symbols. To quantify this discrepancy we calculate the percentages of the \sk values lying outside the $\ebsk\pm\sigma^2_{\ebsk}$ range, $p=15.3\%$ for $M=97$ and $p=22.2\%$ for $M=9766$, which show that, from a practical perspective, the analytical expression given by Eqn. \ref{gauss_varSK} provides a non-standard fluctuation interval that has a larger confidence level of not being crossed by a random sample than a standard $\pm1\sigma$ interval that, in the case of a normal distribution, would be expected to leave $31.73\%$ of the sample points scattered outside its bounds. We note, however, that the percentages of \sk values lying outside the $\langle\esk\rangle(1\pm\epsilon)$ SDEV ranges, $p=26.9\%$ for $M=97$ and $p=31.5\%$ for $M=9766$, do also correspond to higher than standard confidence levels. This behavior is consistent with the positive skewness of the \sk estimator PDF, which proved to be a non-negligible effect when $1\pm3\sigma_{\esk}$ RFI detection thresholds were experimentally tested in the first implementation of an \sk spectrometer \citep{sks}.

Having available only the biased approximations of the first two moments of the true \sk distribution associated with a Gaussian transient, \bsk and $\sigma_{\ebsk}^2$, we can obtain neither a Pearson Type IV PDF analytical approximation, nor an alternative three--moment based Pearson Type III approximation that could be accurate enough for practical applications \citep{gsk}. Nevertheless, for the practical purpose of estimating a particular pair $\{\delta, \rho\}$ by means of \sk measurements, one can instead use post facto Monte-Carlo simulations such those illustrated here to estimate the true confidence level corresponding to the particular $\ebsk\pm\sigma_{\ebsk}$ realization. However, to facilitate  the use of the mathematically convenient propagation of errors formalism for translating the \sk statistical fluctuations estimated by means of Monte-Carlo simulations into formal standard deviations $\{\sigma_{\delta}, \sigma_{\rho}\}$, we propose the use of an empirical standard-deviation (SDEV) correction factor, $\gamma$, which would produce an equivalent SDEV range $\ebsk\pm\gamma\sigma_{\ebsk}$ that would leave outside its bounds $31.73\%$ of the simulated \sk random variables.  Fig. \ref{gauss_mc} indicates such SDEV corrections (dot--dashed lines) that, for $\gamma=0.7013$ and $\gamma=0.8093$, result in leaving outside their corresponding ranges $p=31.7\%$ and $p=32.1\%$ of the scattered random variables, for $M=97$ and $M=9766$, respectively. We note that, as $M$ increases, not only the bias of the \bsk approximation becomes practically negligible, but also the differences between the $\langle\esk\rangle(1\pm\epsilon)$ , $\ebsk\pm\sigma_{\ebsk}$, and $\ebsk\pm\gamma\sigma_{\ebsk}$ ranges, which indicate that, as the \sk distribution approaches normality, the \bsk and $\sigma_{\ebsk}^2$ approximations become asymptotically unbiased. We thus conclude that Eqns. \ref{gauss_sk_approx} and \ref{gauss_varSK} provide approximations suitable for accurate estimations of the characteristics of Gaussian transients.

To illustrate the expected performance of the \sk estimator in detecting Gaussian transients, we show in Fig. \ref{gauss_mc}a,b (horizontal lines) the Pearson Type IV asymmetric detection thresholds  \citep{sk} corresponding to the standard $0.13499\%$ PFA on each side of the unity \sk expectation, i.e. $1_{-0.439}^{+0.903}$ for $M=97$, and $1_{-0.058}^{+0.063}$ for $M=9766$, and in Fig. \ref{gauss_mc}c we display the corresponding percentages of detected transients as function of their duty-cycles (thin line and tick line, respectively). Fig.\ref{gauss_mc}c reveals that, for a given SNR, the \sk detection performance may significantly vary with the transient duty-cycle, as the direct result of the duty-cycle dependence of the \sk estimator and its statistical fluctuations, as seen in panels (a) and (b). However, for large accumulation lengths, as is the case of $M=9766$, the detection performance is a flat $100\%$, except for narrow ranges close to the both ends of the $[0-100]\%$ duty-cycle interval.

\begin{figure}
 \centerline{\includegraphics[width=1\columnwidth]{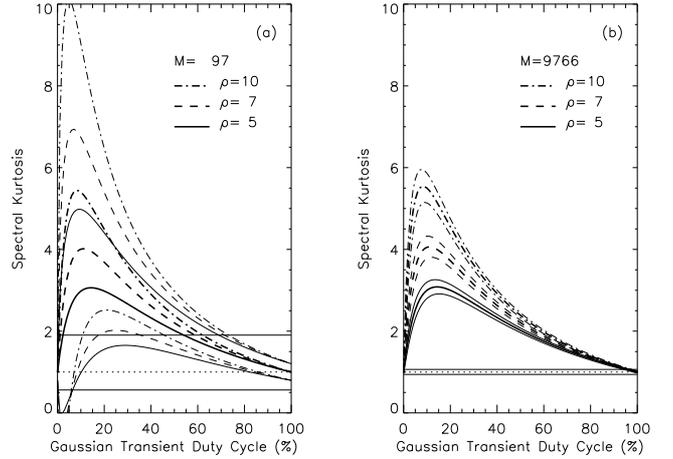}}
 \caption{Gaussian transient duty-cycle variation of the \bsk estimator (thick lines) and its expected $\ebsk\pm\sigma^2_{\ebsk}$ fluctuations (thin lines) for $M=97$ (panel a) and $M=9766$ (panel b), and three selected signal-to-noise ratios, $\rho=5,7,10$, (solid, dashed, and dot--dashed lines, respectively)}
 \label{gauss_overlap}
 \end{figure}

The overall transient detection performance of the \sk estimator is expected to increase as the signal-to-noise ratio increases, as indicated by Fig. \ref{gauss_overlap}, which displays the duty-cycle variation of the \bsk estimator (thick lines), and its expected $\ebsk\pm\sigma^2_{\ebsk}$ fluctuations (thin lines), for $M=97$ (panel a), $M=9766$ (panel b), and three selected signal-to-noise ratios, $\rho=5,7,10$, (solid, dashed, and dot--dashed lines, respectively). However, as shown by \ref{gauss_overlap}a, the $\ebsk\pm\sigma^2_{\ebsk}$  ranges corresponding to different signal-to-noise ratios may overlap, which could result in large uncertainties of the SNR and duty-cycle estimates obtained from \sk measurements. Although, as suggested by \ref{gauss_overlap}b, such uncertainties are expected to decrease as the accumulation length $M$ increases, a more quantitative investigation is called for, which is presented in the next section.

\subsection{\sk measurements of Gaussian transients}\label{gauss_measurements}
To demonstrate the ability of an \sk spectrometer to measure Gaussian transients, we consider the favorable case of a Gaussian transient fully contained within the bounds of a single accumulation characterized by the accumulated power $S_1(t_i)$ and proceeded by a transient--free accumulation $S_1(t_{i-1})$. Based on such two consecutive power measurements,  one may define the apparent signal-to-noise ratio
\begin{equation}
\label{eta}
\eta=\frac{S_1(t_i)}{S_1(t_{i-1})}-1,
\end{equation}
which simply relates to the true SNR and duty-cycle of the transient as $\eta=\delta\rho$.
Hence, from Equation \ref{gauss_sk_approx}, immediately follows
\begin{eqnarray}
\label{delta}
\delta&=&\frac{2\eta^2}{(1 + \eta)^2\esk+(\eta-2)\eta-1 }\\\nonumber
\rho&=&\frac{\eta}{\delta},
\end{eqnarray}
where all magnitudes entering the expressions of $\delta$ and, subsequently, $\rho$ depend only on quantities directly measured by the \sk spectrometer.

Using the general propagation-error formula \citep{Bevington}
\begin{equation}
\sigma_{f(x_i)}^2=\sum_i[\partial_{x_i}f(x_i)]^2\sigma_{x_i}^2,
\end{equation}
the experimental uncertainties of such estimates can be expressed as
\begin{eqnarray}
\label{sigma-delta-rho}
\\\nonumber
\sigma_{\delta}^2
&=&\frac{(1 + \eta)^2}{ 4 \eta^6}
\Big[ \eta^2 (1 + \eta)^2 \sigma_{\widehat{SK}}^2 + 4 (\widehat{SK} - 1)^2 \sigma_{\eta}^2\Big] \delta^4;\\
\sigma_{\rho}^2&=&\Big(\frac{\sigma_\delta^2}{\delta^2}+\frac{\sigma_\eta^2}{\eta^2}\Big)\rho^2,\nonumber
\end{eqnarray}
with $\sigma_{\esk}^2$ being provided by Equation \ref{gauss_varSK}, (optionally scaled by an SDEV correction factor $\gamma$ obtained from Monte-Carlo simulations involving the estimated parameters $\delta$ and $\rho$), and
\begin{eqnarray}
\label{sigma-eta}
\sigma_\eta^2=\Big[\frac{\sigma_{S_1}^2(t_{i})}{S_1^2(t_{i})}+\frac{\sigma_{S_1}^2(t_{i-1})}{S_1^2(t_{i-1})}\Big](1+\eta)^2,
\end{eqnarray}
where the variances $\sigma^2_{S_1}$ can be straightforwardly expressed in terms of the expectations $E[S_1]$ and $E[S_1^2]$, which are provided by Equation \ref{S1mom} as
\begin{eqnarray}
\label{gauss_s1_moments}
E[S_1(M,\mu)]&=&(1+\delta\rho)M\mu\\\nonumber
E[S_1^2(M,\mu)]&=&[1+(2+\rho)\delta\rho+(1+\delta\rho)^2M]M\mu^2.
\end{eqnarray}

Hence, taking into account that $\sigma^2_{S_1}=E[S_1^2]-E[S_1]^2$ and $\rho(t_{i-1})=0$, Equation \ref{sigma-eta} reduces to
\begin{eqnarray}
\label{var_etagauss}
\sigma_{\eta}^2=\frac{1}{2M}\big[4 + 7\eta  + \eta^3 +\eta (\eta-1)^2 \esk \big].
\end{eqnarray}
Therefore, in the case of an observed Gaussian transient, Equations \ref{SK}, \ref{gauss_varSK}--\ref{sigma-delta-rho} and \ref{var_etagauss} provide the theoretical means for estimating its signal-to-noise ratio and duty-cycle, as well as the corresponding statistical uncertainties, in terms of the $S_1$ and $S_2$ measurements provided by the \sk spectrometer.

 \begin{figure}
 \centerline{\includegraphics[width=1.\columnwidth]{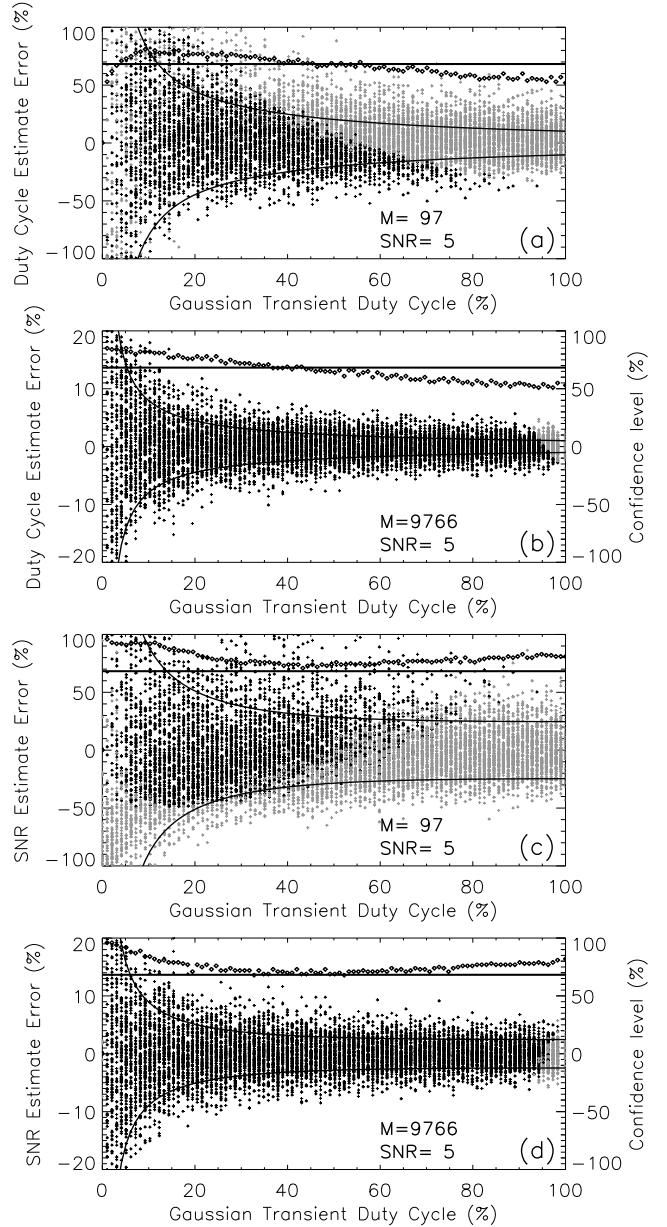}}
 \caption{ Duty cycle distributions of the relative errors of the duty-cycle estimates (panels a and b, $M=97$ and $M=9766$, respectively) and SNR estimates (panels c and d, $M=97$ and $M=9766$, respectively) obtained from the same set of Gaussian transient simulations used to generate the \sk distributions shown in Figure~\ref{gauss_mc}. The estimates obtained from transients flagged by the $0.13499\%$ PFA \sk detection thresholds are indicated by black plus symbols, and the estimates obtained from not flagged \sk simulations are shown as grey symbols. The pairs of solid curves in each panel indicate the range of the expected standard--equivalent $\pm1\sigma$ statistical fluctuations, as derived from Equation {sigma-delta-rho} in which the known true values $\delta_{true}$ and $\rho_{true}$ have been entered, and the correction factors $\gamma=0.7013$ and  $\gamma=0.8093$, for $M=97$ and $M=9766$, respectively, have been applied to the corresponding $\widehat{SK}(\delta_{true},\rho_{true})$ expected fluctuations, as provided by Equation \ref{gauss_varSK}. The square symbols shown in each panel indicate the true confidence levels of the $\pm1\sigma$ intervals, which are compared with a standard $68.27\%$ confidence level (horizontal lines). Due to much smaller fluctuations affecting the $M=9766$ estimates, different scales are used to display on the same plots the relative errors and confidence levels in panels (c) and (d).} \label{gauss_estimate}
 \end{figure}

To investigate the accuracy of the estimations provided by this method, we display in Figure~\ref{gauss_estimate} the results obtained in the case of the Monte-Carlo simulations used to generate the \sk distributions shown in Figure~\ref{gauss_mc}.

Panels \ref{gauss_estimate}a and \ref{gauss_estimate}b display (plus symbols), for $M=97$ and $M=9760$, respectively, the relative errors of the duty-cycle estimates versus the true duty cycles of the simulated transients. The pairs of solid curves shown in both panels indicate the expected $\pm\sigma_{\delta}/\delta_{true}$ range, as inferred from Equation \ref{sigma-delta-rho}, which appear to be in good qualitative agreement with the trend of the observed fluctuations of the duty-cycle estimates. To help quantify this comparison, the square symbols indicate, for each simulated duty-cycle, the percentages of duty-cycle estimates laying within these ranges, i.e. the confidence levels of the $\delta\pm\sigma_{\delta}$ intervals. This comparison show that, for all duty cycles, these confidence levels remain close to the $68.27\%$ standard deviation confidence level we intended to achieve by applying the SDEV correction factors to the $\sigma_{\ebsk}^2$ variances estimated by Equation \ref{gauss_varSK}, i.e. $\gamma=0.7013$ for $M=97$, and  $\gamma=0.8093$ for $M=9766$. However, the duty-cycle trend they follow, indicate a systematic overestimation of the true standard--equivalent $\sigma_{\delta}$ fluctuations for $\delta<\simeq 50\%$, and a systematic underestimation for larger duty cycles. Similarly, panels \ref{gauss_estimate}c and \ref{gauss_estimate}d display the relative errors of the SNR estimates, their expected $\pm\sigma_{\rho}/\rho_{true}$ intervals, as inferred from Equation \ref{sigma-delta-rho}, and the confidence levels of these intervals, which appear to have a duty-cycle dependence that is systematically higher than a standard $68.27\%$ confidence level. Therefore, for any duty-cycle, Equation \ref{sigma-delta-rho} systematically overestimate the true standard--equivalent fluctuations of the $\rho$ estimates.

Figure~\ref{gauss_estimate} demonstrates that the statistical fluctuations of both $\delta$ and $\rho$ estimates, which appear to be symmetrically distributed around the true parameter values, significantly decrease as the duty-cycle increases from zero to about $50\%$, and continue to decrease at a more slower pace, as the duty-cycle approaches $100\%$. Remarkably, these fluctuations decrease by one order of magnitude as the accumulation length increases from $M=97$ to $M=9766$.

However, we note that all of the above conclusions are based on the estimates obtained from all simulated Gaussian transients, disregarding whether or not they were actually flagged as such by the $0.13499\%$ PFA detection thresholds shown in Figure~\ref{gauss_mc}. The make such distinction, we use black symbols to display the estimates obtained from \sk flagged measurements, and grey symbols for those obtained from those transients that would remain undetected in a real--life experiment. The ratio of black symbols to $n=1000$, which is the total number of simulated transients having the same duty-cycle, follows, in each case, the duty-cycle dependence of the \sk detection performance shown in Figure~\ref{gauss_mc}c. From this perspective, we have to make the cautionary note that, although any individual \sk measurement may be affected by practically small statistical uncertainties, the sample means of the $\delta$ and $\rho$ estimates corresponding to a group of Gaussian transients characterized by the same true SNR and duty-cycle, may be statistically biased.  Nevertheless, as demonstrated by the panels corresponding to $M=9766$, this detection--induced statistical bias may be significantly reduced by increasing the accumulation length, or, at the cost of larger probabilities of false alarm, by lowering the detection thresholds for short accumulation lengths.

We thus conclude that the results displayed in Figure~\ref{gauss_estimate} clearly demonstrate the ability of the \sk--based measurement method presented in this section to provide estimates of the true parameters that, for sufficiently large accumulation lengths,  may be affected by standard--equivalent statistical fluctuations not larger than a few percent.

\section{Statistics of coherent transients}\label{rfi}
In this section we analyze the statistical properties of the \sk estimator in the case of a coherent transient time domain signal mixed with a quasi-stationary Gaussian background. In \S\ref{stationary_rfi_meanvar} we obtain a generally valid expression for the \sk estimator that, in the limit $M\gg1$, reduces to the $M\gg1$ approximation of an expression previously reported by \citet{rfi}, and we also obtain a first order approximation for the variance of this estimator. In \S\ref{transient_rfi_meanvar} we generalize these results and provide analytical expressions for the \sk estimator and its variance in the case of a transient coherent signal mixed with a Gaussian time domain background.  We validate these analytical expectations by means of numerical simulations.
\subsection{Quasi-stationary coherent signals}
\label{stationary_rfi_meanvar}
To determine the statistical properties of the \sk estimator associated with coherent signals mixed with Gaussian background, we follow the same framework employed in \S\ref{gauss_transients} for the case of Gaussian transients, with the only difference being that the underlaying probability distribution function of the raw FFT--derived PSD estimate $P$ is in this case given by \citep{SineRFI}
\begin{eqnarray}\label{SineRFI}
pdf(P)=\frac{1}{2\sigma^2}\exp\Big(-\frac{P+A^2}{2\sigma^2}\Big)I_0\Big(\frac{\sqrt{P}A}{\sigma^2}\Big),
\end{eqnarray}
where $\sigma^2$ represents the variance of a time domain Gaussian background, $A$ is the amplitude of a mixed time-domain sinusoidal signal, and $I_\alpha$ is the modified Bessel function of first kind.

Taking in consideration that, at each frequency bin, the mean spectral power of the time domain Gaussian background relates to the time domain variance $\sigma^2$ as $\mu=2\sigma^2$ \citep{rfi}, and defining the signal-to-noise ratio of the mixed coherent signal as $\rho=A^2/2\sigma^2$, Eqn. \ref{SineRFI} may be rewritten in terms of the normalized random variable $x\equiv P/\sigma^2=2P/\mu$ as a non-central chi-square distribution with $k=2$ degrees of freedom and non-centrality parameter $\lambda=2\rho$, i.e. $\chi^2_{pdf}(x,2,2\rho)$, where
\begin{eqnarray}\label{chisqrpdf}
\chi^2_{pdf}(x,k,\lambda)=\frac{1}{2}\exp\Big(-\frac{x+\lambda}{2}\Big)I_{k/2-1}\Big(\sqrt{x \lambda}\Big).
\end{eqnarray}
Using the linearity property of the expectation operator, and assuming statistical independence of the time series samples, the PDF given by Eqn. \ref{chisqrpdf} provides
\begin{eqnarray}\label{rfi_s1s2}
E[S_1]=M(1+\rho)\mu;\;\;E[S_2]=M (2 + 4\rho+\rho^2)\mu^2.
\end{eqnarray}
To compute the expectation $E(S_1^2)$, which is needed to evaluate the expectation of the \sk estimator, we derive the probability distribution of a sum of $M$ independent $\chi^2_{pdf}(x,2,2\rho)$--distributed random variables, which is $\chi^2_{pdf}(x,2 M,2\rho M)$, from which we get
\begin{equation}
\label{rfi_s12}
E[S_1^2]= [1 + 2 \rho + M (1 + \rho)^2]M \mu^2.
\end{equation}
Since \citep{sk},
\begin{eqnarray}
cov\Big(\frac{S_2}{S_1^2},S_1^2\Big)=\frac{1}{M^2}\Big{[}\frac{2}{\mu_1'}(\mu_3'-\mu_1'\mu_2')-\frac{4\mu_2'}{\mu_1'^2}(\mu_2'-\mu_1'^2)\Big{]},
\end{eqnarray}
where, $\mu_1'\equiv\mu$ and $\mu_k$ are the raw $\chi^2_{pdf}$ moments of order $k$, we have
\begin{equation}
\label{covrfi}
cov\Big(\frac{S_2}{S_1^2},S_1^2\Big)=-\frac{4 \mu^2 \rho^2}{M^2 (1 + \rho)^2},
\end{equation}
which, combined with the identity
\begin{equation}
cov\Big(\frac{S_2}{S_1^2},S_1^2\Big)=E[S_2]-E\Big[\frac{S_2}{S_1^2}\Big]E[S_1^2],
\end{equation}
leads to the unbiased expectation
\begin{equation}
\label{rfiratio}
E\Big[\frac{S_2}{S_1^2}\Big]=\frac{E[S_2]}{E[S_1^2]}+\frac{4 \mu^2 \rho^2}{M^2 (1 + \rho)^2}\frac{1}{E[S_1^2]}.
\end{equation}
Hence, combining Eqns. \ref{rfi_s1s2}, \ref{rfi_s12}, and \ref{rfiratio}, immediately follows
\begin{eqnarray}\label{stationary_rfi_sk}
E\Big[\esk\Big]&=&1 - \frac{M \rho^2}{1 + 2 \rho + M (1 + \rho)^2}+O\Big(\frac{1}{M^3}\Big)\\\nonumber
&\simeq& 1 - \frac{\rho^2}{(1 + \rho)^2},
\end{eqnarray}
where the $M\gg1$ approximation of the \sk expectation is identical with the spectral variability of the parent population of the PSD estimate, $\sigma^2_P/\mu_P^2$ \citep{rfi}. This proves that the \sk estimator defined by Eqn. \ref{SK}, which is an unbiased estimator of the spectral variability of a Gaussian time domain signal \citep{sk}, is also a biased estimator of the spectral variability of a coherent signal mixed with a Gaussian background.

We note that, for $\rho=0$, Eqn. \ref{stationary_rfi_sk} reduce to unity, as expected for a transient-free Gaussian background, while it decreases from unity toward zero as fast as $2/\rho$, as $\rho$ increases.

To obtain an analytical approximation for the variance of the \sk estimator similar to Eqn. \ref{gauss_varSK}, in addition to $E[S_1^2]$ and $E[S_2]$, one would not only need to obtain an analytical expression for the expectations $E[S_1^n]$, $(n=\overline{1,4})$, which can be exactly computed from the known $\chi^2_{pdf}(x,2,2\rho)$ distribution, but also the expectation $E[S_2^2]$, which must be computed from the parent distribution of the $S_2$ random variable, for which, so far, we were not able to find a closed-form analytical expression. Instead, we compute a first order approximation of $\sigma^2_{\widehat{SK}}$ that is generally valid for any $M$ and for any probability distribution of the raw PSD estimates \citep{rfi}, which can be written in terms of the expectations $E[x]$ as
\begin{eqnarray}\label{varsk_firstorder}
\sigma_{\widehat{SK}}^2\approx &&\Big(\frac{M+1}{M-1}\Big)^2\frac{1}{M}\Big(\frac{E[x^4]-E[x^2]^2}{E[x]^4}\\\nonumber
&&-\frac{4E[x^2]E[x^3]}{E[x]^5}+\frac{4E[x^2)^3}{E[x]^6}\Big).
\end{eqnarray}

For the particular case of $x$ being distributed according to $\chi^2_{pdf}(x,2,2\rho)$, Eqn. \ref{varsk_firstorder} leads to
\begin{equation}\label{stationary_rfi_varsk}
\sigma_{\widehat{SK}}^2\approx
\frac{4(1 + 6 \rho + 10 \rho^2 + 8 \rho^3 + 2 \rho^4)}{M(1+\rho)^6},
\end{equation}
where, in addition to the first order approximation in terms of $\rho$,  we have dropped the contribution of $(M+1)^2/(M-1)^2$, which becomes negligible for $M\gg1$.

We note that for $\rho=0$, as expected, Eqn. \ref{stationary_rfi_varsk} reduce to the $M\gg1$ approximation ($4/M$) of the variance of the \sk estimator associated with a quasi-stationary time domain Gaussian signal, while it vanishes as fast as $8/M\rho^2$, when $\rho$ goes to infinity.
\subsection{Coherent transient signals}\label{transient_rfi_meanvar}
To investigate how the \sk estimator is expected to deviate from unity in the case of coherent time domain signal lasting shorter than the accumulation time, we employ the same framework as in \S\ref{gauss_transients}, with the only difference of replacing the underlaying gamma statistical distribution characteristic to a Gaussian PSD estimate \citep{rfi,sk,gsk}, with the $\chi_{pdf}^2(x,2,\rho)$ distribution (Eqn. \ref{chisqrpdf}) describing the statistical properties of the PSD estimates corresponding to a coherent time domain signal characterized by a signal-to-noise ratio $\rho$ relative to a quasi-stationary Gaussian time domain background. This approach straightforwardly leads to the biased expectation
\begin{eqnarray}
\label{rfi_sk_full}
\ebsk=1 + \frac{(1 - 2 \delta)\delta \rho^2  M^2 +
    \delta \rho^2M}{(1 + \delta \rho)^2 M^2 - \delta^2\rho^2  M - (1 + 2 \delta \rho)},
\end{eqnarray}
which, for accumulation lengths $M\gg1$,  reduces to
\begin{eqnarray}
\label{rfi_sk_approx}
\ebsk\simeq 1 + \frac{(1 - 2 \delta) \delta \rho^2}{(1 + \delta \rho)^2}.
\end{eqnarray}
As expected, for $\delta$=0 (no transient signal present), both Eqns. \ref{rfi_sk_full} and \ref{rfi_sk_approx} reduce to unity, while for $\delta=1$ (quasi-stationary coherent signal), they reduce to the full \sk expression and its $(M\gg1)$ approximation provided by Eqn. \ref{stationary_rfi_sk}. However, differently from the case of Gaussian transients, which are exclusively characterized by \sk values larger than unity, the \sk expressions provided by Eqns. \ref{rfi_sk_full} and \ref{rfi_sk_approx} may take values larger than unity for $\delta<0.5$ and smaller than unity for $\delta>0.5$, while crossing the $1\pm2/\sqrt{M}$ interval for duty-cycles close to $50\%$, which is a well-known limitation of the \sk or time domain kurtosis--based RFI detection algorithms \citep{ruf,roo,rfi,sk,gsk,sks}. However, for $\delta=1/(4+\rho)$, Eqn. \ref{rfi_sk_approx} reaches its maximum deviation from unity, $\ebsk=1+\rho^2/(8+4\rho)$, which makes \sk a very a very efficient coherent transient detector. Nevertheless, similar to the case of Gaussian transients addressed in \S\ref{gauss_transients}, an evaluation of the \sk statistical fluctuations is needed to fully asses its performance as a detector, as well as the experimental uncertainties affecting any parameter estimated from \sk measurements.

The absence of closed form analytical expressions for the moments of the sums of squared random variables distributed according to $\chi_{pdf}^2(x,2,\rho)$ , which prevented us from obtaining an exact analytical expression for the variance of the \sk estimator associated with a quasi-stationary coherent signal, also prevents us from obtaining an analytical expression in the case of coherent transients. Moreover, the approach we used to in \S\ref{stationary_rfi_meanvar} to obtain the first order approximation of the \sk variance is not directly applicable in the case of coherent transients due o the fact that $S_1$ and $S_2$ are not sums of random variables drawn from the same parent population. Instead, we provide an approximation that, although may seem based on more or less speculative basis, will be proven to be in agreement with the statistical fluctuations observed in numerical simulations.

Our approach is motivated by the observation that, although the full expressions of the \bsk estimators associated with the Gaussian (Eqn. \ref{gauss_sk_full}) and coherent (Eqn. \ref{rfi_sk_full}) transients are mathematically different, their $(M\gg1)$ approximations (Eqns. \ref{gauss_sk_approx} and \ref{rfi_sk_approx}, respectively) are mathematically equivalent in the sense that the same observed \sk value larger than unity may be either the result of a Gaussian transient characterized by the parameter pair $\{\delta,\rho\}$, with $\delta$ anywhere in the $0-100\%$ range, or, alternatively, the result of a coherent transient with a duty-cycle shorter than $50\%$  characterized by the parameters $\{\delta/2, 2\rho\}$. While, in the absence of additional information, this morphological transformation makes in principle indistinguishable the true physical nature of the observed transients exclusively from one \sk measurement, it offers us enough grounds to speculate that the true variance of the coherent transient \sk estimator might be reasonably approximated by applying the same morphological transformation to the $(M\gg1)$ approximation provided by Eqn. \ref{gauss_varSK}. This leads to the approximation
\begin{eqnarray} \label{rfi_varSK}
&&\sigma^2_{\ebsk}\simeq\frac{1}{2 M(1 + \delta \rho)^6 }
   (8 + 48 \delta \rho + 72 \delta \rho^2 \\\nonumber
   &&- 24 \delta^2 \rho^2 +40 \delta \rho^3 + 48 \delta^2 \rho^3 -96 \delta^3 \rho^3 + 5 \delta \rho^4\\\nonumber
   &&+56 \delta^2 \rho^4-72 \delta^3 \rho^4+10 \delta^2 \rho^5 -8 \delta^3 \rho^5 + \delta^3 \rho^6),
\end{eqnarray}
which, for $\delta=0$, reduces to $4/M$, which is indeed the $ (M\gg1)$ expected \sk variance for a quasi-stationary Gaussian time domain signal, while for $\delta=1$, (the case of a quasi-stationary coherent signal),  it reduces to
\begin{equation} \label{stationary_rfi_varsk_approx}
\sigma^2_{\ebsk}\simeq\frac{8 + 48 \rho + 48 \rho^2 - 8 \rho^3 - 11 \rho^4 +2 \rho^5 + \rho^6}{2 M (1 + \rho)^6},
\end{equation}
which needs to be compared with the non-identical approximation that we analytically derived directly from the true statistical distribution of the PSD samples provided by Eqn. \ref{stationary_rfi_varsk}.

\begin{figure}
 \centerline{\includegraphics[width=1\columnwidth,clip=]{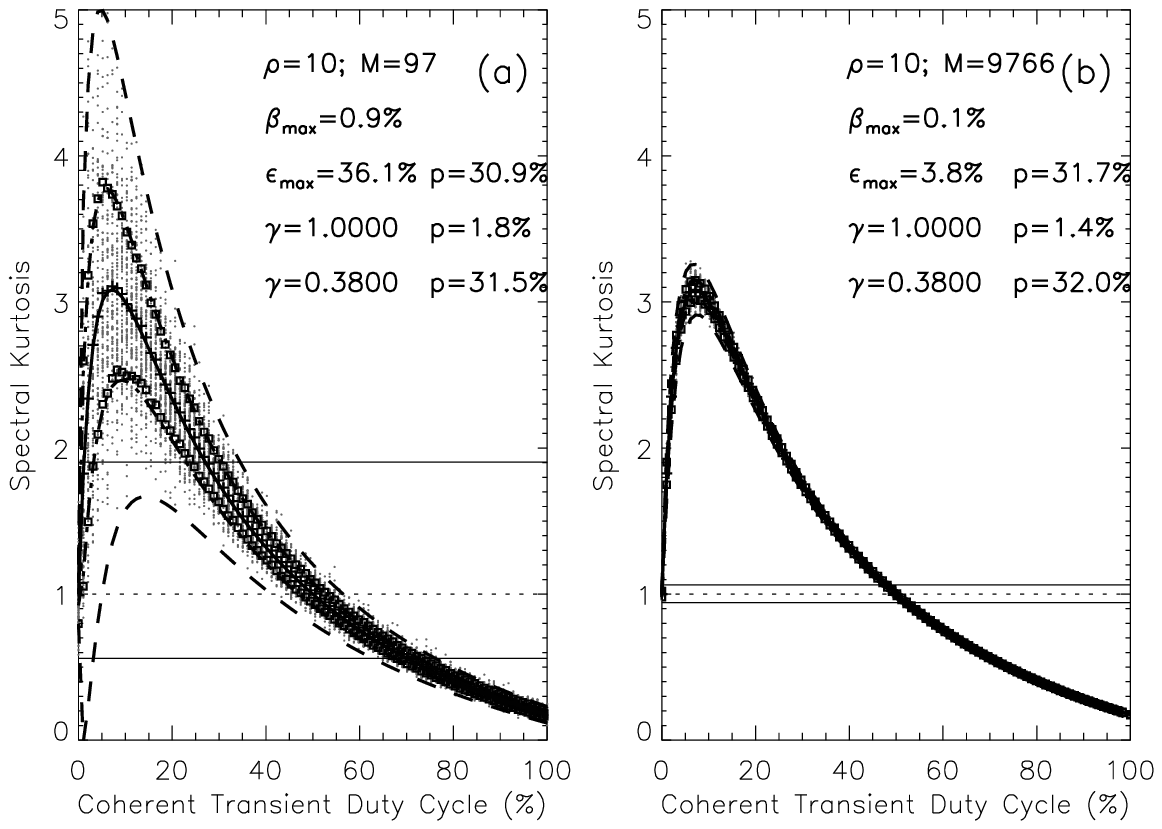}}
 \centerline{\includegraphics[width=1\columnwidth,clip=]{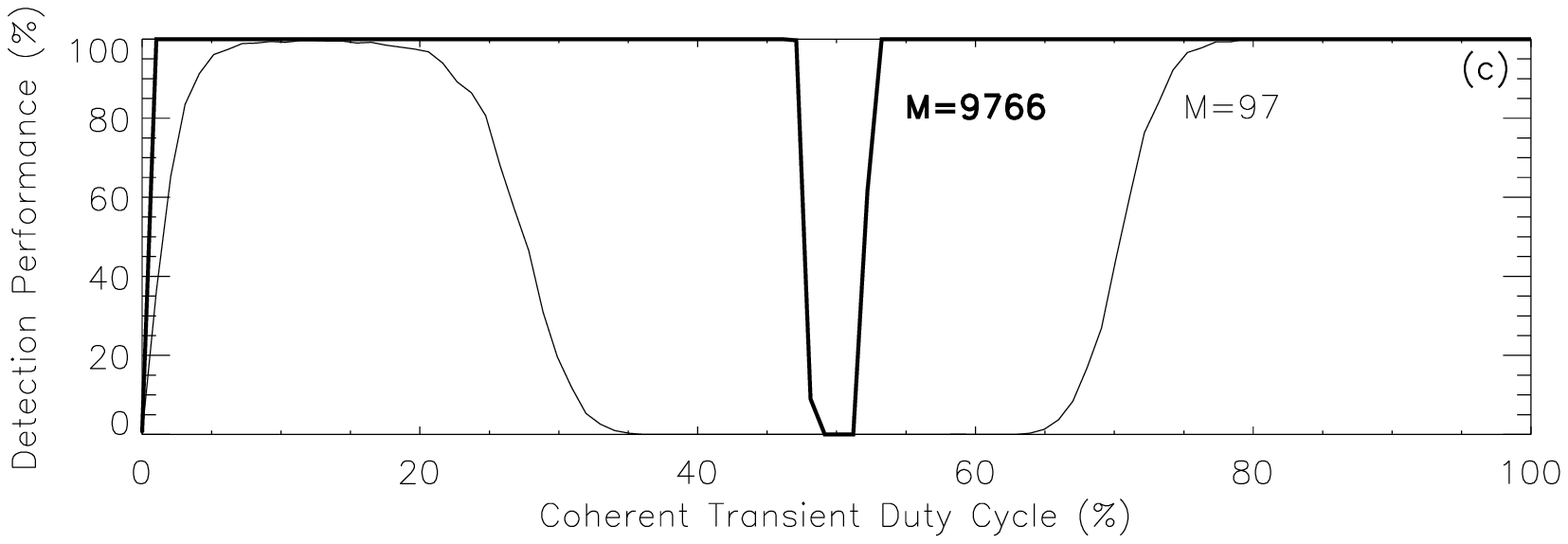}}
  \caption{ \bsk estimator (solid lines) as function of a coherent transient duty-cycle for a signal-to-noise ratio $\rho=10$ and two accumulations lengths, $M=97$ (panel a) and $M=9766$ (panel b). The symmetric $\ebsk\pm\sigma_{\ebsk}$ limits provided by Eqn. \ref{rfi_varSK} are overlaid (dashed lines) on top of the corresponding numerically simulated distributions (point symbols). The SDEV--corrected $68.27\%$ probability ranges,  $\ebsk\pm\gamma\sigma_{\ebsk}$, are indicated by the dot-dashed lines. The square symbols indicate the numerically sample SDEV ranges around the mean of the simulated \sk distribution (plus symbols).  The Pearson Type IV asymmetric detection thresholds, which correspond to standard $0.13499\%$ probabilities of false alarm on each side of the unity \sk expectation if no transient emission was present, are indicated by horizontal lines. The SDEV correction factors $\gamma$, the maximum values reached by the duty-cycle dependent relative standard deviations $\epsilon_{max}$, and the corresponding maximum relative bias of the \sk estimator, $\beta_{max}$, are indicated in each figure inset.}\label{rfi_mc}
 \end{figure}

This comparison reveals that, while Eqn. \ref{stationary_rfi_varsk_approx} also reduces to $4/M$ for $\rho=0$, unlike Eqn. \ref{stationary_rfi_varsk}, it does not completely vanishes as $\rho$ goes to infinity. Instead, it approaches as fast as $1/2M-2/M \rho$ a residual value of $1/2M$ that practically vanishes for large accumulation lengths $M$. Based on this comparison, we find the semi-analytical approximation provided by Eqn. \ref{rfi_varSK} suitable for practical application, especially in the light of analysis illustrated in Fig. \ref{gauss_mc}, which indicated the need of a numerical SDEV correction of the $\sigma_{\ebsk}$ analytical expression.

In Fig. \ref{rfi_mc} we present a similar analysis for the purpose of validating the analytical expressions obtained in this section. Remarkably, when compared with the sample mean of the Monte-Carlo simulations (plus symbols), the inaccuracy of the \bsk approximation (solid red lines) appears to be negligible even for relatively short accumulation lengths. Nevertheless, for both values of $M$, the $\ebsk\pm\sigma_{\ebsk}$  ranges (dashed lines) appear to largely overestimate the sample standard deviation (square symbols). In this simulation, we find that the same SDEV correction factor $\gamma=0.38$ is needed to be applied for both values of $M$ to assure that $68.27\%$ of the simulated samples are scattered within the $\ebsk\pm\gamma\sigma_{\ebsk}$  ranges indicated by the dot--dotted lines.

 \begin{figure}
 \centerline{\includegraphics[width=1\columnwidth]{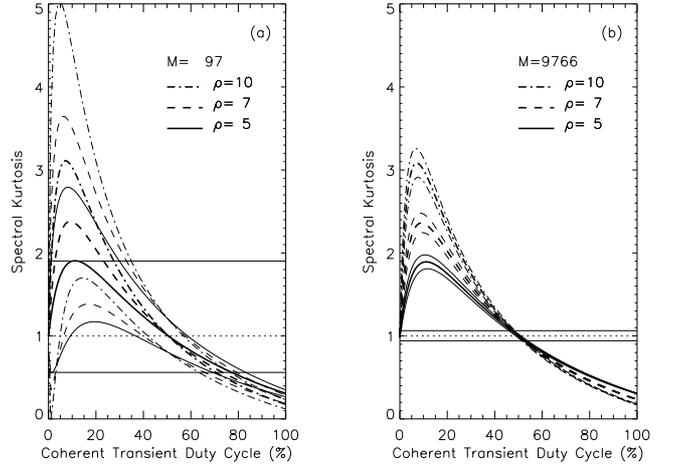}}
 \caption{Coherent transient duty-cycle variation of the \bsk estimator (thick lines) and its expected $\ebsk\pm\sigma^2_{\ebsk}$ fluctuations (thin lines) for $M=97$ (panel a) and $M=9766$ (panel b), and three selected signal-to-noise ratios, $\rho=5,7,10$, (solid, dashed, and dot--dashed lines, respectively)} \label{rfi_overlap}
 \label{coherent_overlap}
 \end{figure}

The same as in the case of Gaussian transients, the \sk performance in detecting coherent transients, which is illustrated Figure~\ref{rfi_mc}c, improves as the accumulation length increases, reaching a flat $100\%$ for all duty-cycles except narrow ranges at both ends of the interval, as well as around the $50\%$ duty-cycle mark. Fig. \ref{gauss_overlap} completes this detection performance analysis by showing that the coherent transient detection performance of the \sk estimator increases as the signal-to-noise ratio increases. However, as shown by \ref{gauss_overlap}a, the $\ebsk\pm\sigma^2_{\ebsk}$  ranges corresponding to different  signal-to-noise ratios may overlap, which could result in large uncertainties of the SNR and duty-cycle estimates obtained from \sk measurements. This aspect is quantitatively investigated in the next section.

\subsection{\sk measurements of coherent transients}\label{coherent_measurements}
\begin{figure}
 \centerline{\includegraphics[width=1.\columnwidth]{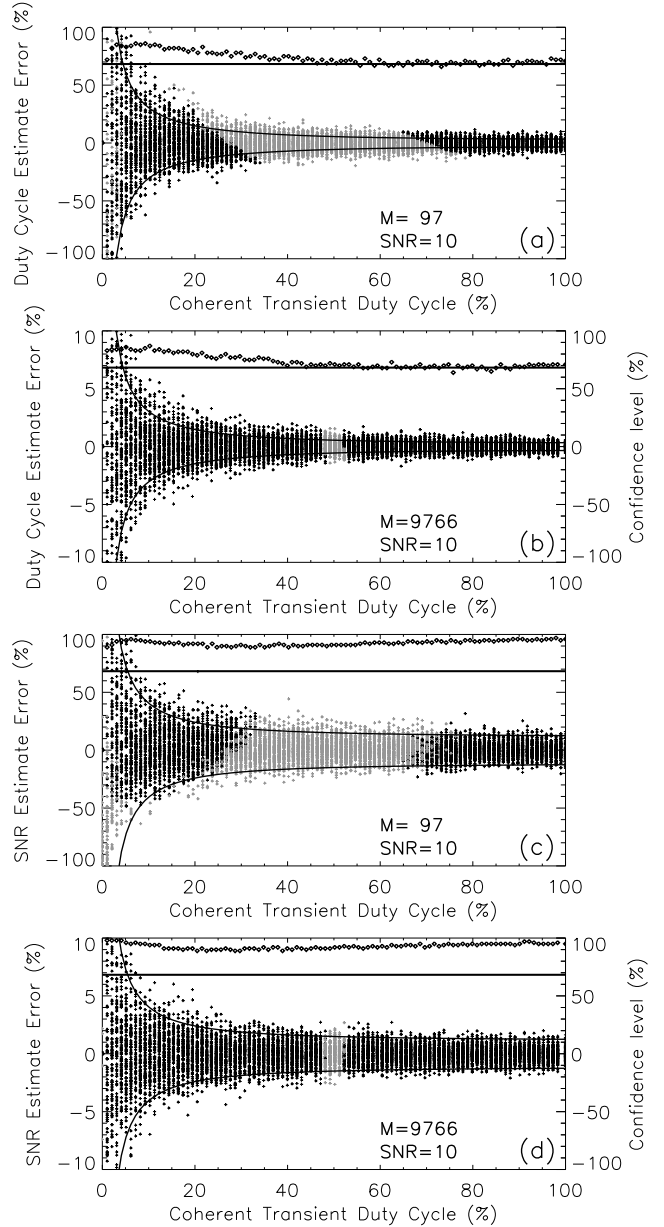}}
 \caption{ Duty cycle distributions of the relative errors of the duty-cycle estimates (panels a and b, $M=97$ and $M=9766$, respectively) and SNR estimates (panels c and d, $M=97$ and $M=9766$, respectively) obtained from the same set of coherent transient simulations used to generate the \sk distributions shown in Figure~\ref{rfi_mc}. The estimates obtained from transients flagged by the $0.13499\%$ PFA \sk detection thresholds are indicated by black plus symbols, and the estimates obtained from not flagged \sk simulations are shown as grey symbols. The pairs of solid curves in each panel indicate the range of the expected standard--equivalent $\pm1\sigma$ statistical fluctuations computed based on the known true values $\delta_{true}$ and $\rho_{true}$. The same correction factor $\gamma=0.38$ has been applied for both $M=97$ and $M=9766$ to the corresponding $\widehat{SK}(\delta_{true},\rho_{true})$ expected fluctuations, as provided by Equation \ref{rfi_varSK}. The square symbols shown in each panel indicate the true confidence levels of the $\pm1\sigma$ intervals, which are compared with a standard $68.27\%$ confidence level (horizontal lines). Due to much smaller fluctuations affecting the $M=9766$ estimates, different scales are used to display on the same plots the relative errors and confidence levels in panels (c) and (d).} \label{coherent_estimate}
 \end{figure}

Following the same approach as in \S\ref{gauss_measurements}, and taking in consideration that, in the case of a coherent transients mixed with a Gaussian, the variance $\sigma_{S_1}^2= E(S_1^2)-E(S_1)^2$ can be expressed in terms of the expectations provided by Equations \ref{rfi_s1s2} and \ref{rfi_s12}, and $\sigma_{\esk}^2=\gamma\sigma_{\ebsk}^2$ is provided by Equation \ref{rfi_varSK}, the steps leading to the SNR and duty-cycle estimates, and their corresponding statistical uncertainties, is fully described by the following sequence of equations, which ultimately depend only on the directly measured magnitudes $S_1(t_i)$, $S_2(t_i)$, and $S_1(t_{i-1})$:
\begin{eqnarray}
\label{rfi_workflow}
\label{delta}
\esk&=&\frac{M+1}{M-1}\Big[\frac{M S_2(t_i)}{S_1^2(t_i)}-1\Big];\\\nonumber
\eta&=&\frac{S_1(t_i)}{S_1(t_{i-1})}-1;\\\nonumber
\delta&=&\frac{\eta^2}{(1 + \eta)^2\esk+(\eta-2)\eta-1 };\\\nonumber
\rho&=&\frac{\eta}{\delta};\\\nonumber
\sigma_\eta^2&=&\frac{1}{M}(2 + 4\eta +\eta^2);\\\nonumber
\sigma_{\delta}^2
&=&\frac{(1+\eta)^2}{\eta^6}
\Big[\eta^2(1+\eta)^2\sigma_{\widehat{SK}}^2+4(\widehat{SK}-1)^2\sigma_{\eta}^2\Big]\delta^4;\\\nonumber
\sigma_{\rho}^2&=&\Big(\frac{\sigma_\delta^2}{\delta^2}+\frac{\sigma_\eta^2}{\eta^2}\Big)\rho^2.
\end{eqnarray}
Figure~\ref{coherent_estimate}, which has the same layout as  Figure~\ref{gauss_estimate}, illustrates the performance of the estimations provided by Equation \ref{rfi_workflow} in the case of the coherent transient simulations characterized by the \sk distributions shown in Figure~\ref{rfi_mc}.  Figure~\ref{coherent_estimate} demonstrate that the workflow described by Equation \ref{rfi_workflow} may provide SNR and duty-cycle estimates that, even for relatively short accumulation lengths, are affected by statistical uncertainties that, for duty-cycles larger than about $20\%$ do not exceed a few percent. We also find that the confidence levels of the $\delta\pm\sigma_{\delta}$ intervals, also provided Equation \ref{rfi_workflow}, are consistent with a standard $68.27$ confidence level for most of the duty-cycle interval, while the confidence levels of the $\rho\pm\sigma_{\rho}$ intervals are systematically higher. We thus conclude that \sk--based measurement method presented in this section has a level of accuracy is suitable for practical applications.
\section{\sk discrimination of underlaying transient statistics}
In the previous sections we demonstrated the ability of an \sk spectrometer to detect and measure two special categories of spectral transients  mixed with a Gaussian time domain background. However, this performance analysis involved prior knowledge of the true, Gaussian or coherent, statistical nature of the transients. Therefore, the ability of inferring the underlaying transient statistics from \sk measurements has still to be demonstrated. For this purpose, we consider the hypothetical case of two transients, one Gaussian and another coherent, that have the same signal-to-noise ratios and durations, and investigate the variation of their expected \sk estimator as function of various accumulation lengths.

\begin{figure}
 \centerline{\includegraphics[width=1.\columnwidth]{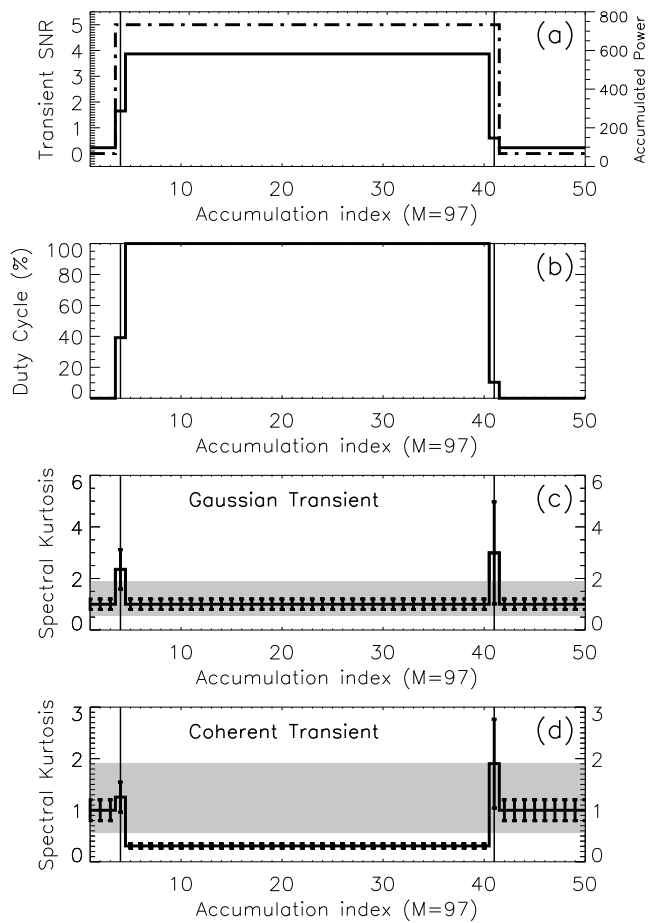}}
 \caption{ Expected \sk discrimination of two transients lasting longer than the accumulation length ($M=97$).  The transients, which have different underplaying statistics,  have the same duration (3540 raw FFT blocks) and SNR ($\rho=5$), and start at the same offset (350  raw FFT blocks) relative to the start of the first accumulation. a) SNR (dot-dotted line) and accumulated power (solid line) as function of the accumulation block index. b) The duty-cycle profile of both transients. \sk (solid line) and $\esk\pm\sigma_{\ebsk}$ (error bars) for the Gaussian and coherent transients are shown in panels (c) and (d), respectively. The range bounded by the $0.13\%$PFA detection thresholds, $[0.56,1.90]$, is indicated by the gray--shaded areas in panels (c) and (d). The accumulation blocks during which the transients start and end are marked by vertical lines in all panels.} \label{monoscale}
 \end{figure}

Figure~\ref{monoscale} presents the result of such analysis for an accumulation length set to $M=97$. To model some particular aspects that may be encountered in a real experiment, both transients were purposely chosen to have a SNR $\rho=5$, a duration $\Delta M=3500$ FFT blocks, longer than the accumulation length, and an offset $\delta M=50$ FFT blocks relative to the start of one of the accumulation blocks. Figures~\ref{monoscale}a and \ref{monoscale}b display the accumulated power and, respectively, the duty-cycle, which both have flat distributions over all but the two accumulation blocks containing the rising and falling edges of the transients. Consequently, as shown in Figure~\ref{monoscale}c, the expected \sk of the Gaussian transient deviates from unity only in the accumulations bins that do not have a duty-cycle equal to $0\%$ or $100\%$. However, due to their relatively large statistical fluctuations, the rising and falling edges of such Gaussian transients may or may not be flagged by the $0.13\%$ PFA detection thresholds, i.e. $[0.56,1.90]$. On the contrary, in Figure~\ref{monoscale}d, all inner accumulations blocks are flagged as unambiguously containing a coherent transient, because the exact $100\%$ duty cycle translates into less than unity \sk values. However, the rising edge of such a coherent transient would escape detection due to its $\sim50\%$ duty-cycle, while its falling edge, which corresponds to a duty-cycle $\sim11\%$, may or may not escape detection due to its relatively large statistical fluctuation, despite an $\esk=1.90$ that happens to be close the maximum value attainable by a coherent transient having $\rho=5$, which is $\delta=1/9=11.11\%$ (Equation \ref {rfi_sk_approx}).

Therefore, the results illustrated by Figure~\ref{monoscale} indicate that, for accumulation lengths shorter than the transient duration, the \sk analysis alone is guaranteed to detect $100\%$ duty-cycle transients, unambiguously recognize their coherent dynamics, and even directly measure their duration with an uncertainty comparable with the integration time. However, a Gaussian transient having the same $100\%$ duty-cycle may entirely escape \sk detection. Moreover, even if both rising and falling edges are detected, they could not be unambiguously attributed to the edges of a a Gaussian transient, since they could be equally attributed to two unrelated transients, of any of the two types, having durations shorter than half of the integration time. Nevertheless, if the existence of such Gaussian transient is alternatively flagged by its accumulated power profile, the \sk analysis may unambiguously determine the  nature of such transient, since only Gaussian transients may have $100\%$ duty-cycles and unity \sk. However, such $S1$--based transient detection scheme, which would necessarily involve arbitrarily defined empirical detection thresholds, would not be as reliable as an \sk--only detection scheme based on exactly known probabilities of false alarm \citep{sk}.

Based on this analysis presented in Figure~\ref{monoscale}, we conclude that an \sk spectrometer may efficiently flag continuous or transient coherent signals longer than its integration time, as well as both Gaussian and coherent transients shorter than its integration time, but without being able to unambiguously discriminate the statistical nature of such short transients. Nevertheless, based on a combined $S_1$ and \sk analysis, the statistical nature of the transients lasting longer than the integration time could be inferred, and thus, their duration and signal-to-noise ratios estimated based on the correct statistical model.

The first experimental validation of such \sk spectrometer capabilities has been provided by \citet{rfi}. Using data recorded by the FST instrument during a solar radio burst, and a software--implemented \sk spectrometer design involving sets of $100$~$\mu$s  contiguous acquisition blocks followed by $20$~ms acquisition gaps,  \citet{rfi} demonstrated the ability of the \sk spectrometer to selectively filter out RFI transients shorter or longer than the integration time, while leaving untouched the microwave spikes of solar origin, which were inferred to have durations longer than the $100$~$\mu$s accumulation time.

However, in a more recent study,  using data obtained with the EST instrument during another solar bursts featuring spiky emission, and an hardware--implemented \sk spectrometer that was designed to integrate $20ms$ contiguous blocks $(M=9766)$, with no acquisition gaps in between, \citet{jgr} demonstrated that the integration blocks containing radio spikes of solar origin were flagged by the \sk $0.13\%$ PFA detection thresholds. Using the analysis framework detailed in \S\ref{gauss_measurements}, \citet{jgr} estimated that the spectral peak of one of the observed solar radio spikes was characterized by a SNR $\rho=2.14\pm0.11$, and a duration $\tau=(8.05\pm0.30)$~ms, which is consistent with theoretical expectations \citep{spike_duration_theory} and previous time--resolved observations of microwave solar radio spikes \citep{spike_duration_observations}.

Nevertheless, to assign a Gaussian statistical model to the observed microwave spikes, \citet{jgr} had to rely on theoretical expectations that microwave spikes of solar emission must have a Gaussian time domain distribution, and to discard the possibility of \sk flagged spikes to represent low duty-cycle local instrumental RFI, hypothesis that was ruled out by serendipitous Very Large Array \citep[VLA,][]{VLA} simultaneous observations of the same spikes, which independently confirmed their genuine solar origin \citep{Science}.

All of the above examples indicate that, if a targeted class of transients is expected to have durations ranging in a certain interval, the accumulation length of an \sk spectrometer may be in principle tuned to an optimal value that would allow intrinsic discrimination of their underlying statistical properties.

\begin{figure}
 \centerline{\includegraphics[width=1.\columnwidth]{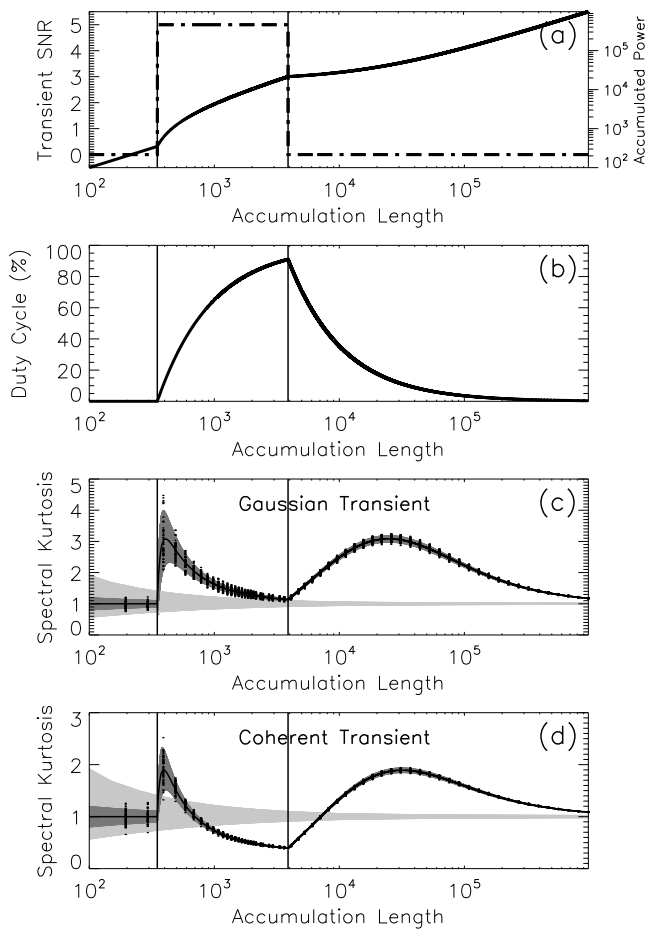}}
 \caption{ Expected \sk profiles as function of a varying accumulation length for the same pair of transients considered in Figure~\ref{monoscale}. The SNR (dot-dotted line) and accumulated power (solid line) profiles are sown in panel (a), and the duty-cycle profile is shown in panel (b). A series of numerically--generated \sk random deviates corresponding to a set of selected integer multiples of the minimum accumulation length, $M=97$, are overlayed (symbols) on the  \sk (solid line) and $\esk\pm\sigma_{\ebsk}$ (dark shaded areas) corresponding to the Gaussian ( panel c) and coherent (panel d) transients.   The range bounded by the $0.13\%$PFA detection thresholds is indicated by the gray--shaded areas in panels (c) and (d). The start and end of the transients are marked by vertical lines in all panels.} \label{multiscale}
 \end{figure}

To explore such possible avenue, Figure~\ref{multiscale} illustrates, for the same hypothetical transients considered in Figure~\ref{monoscale}, the expected \sk profiles obtained by varying the accumulation length in unit steps, from $M=97$, up to a maximum accumulation length several order of magnitude larger. In addition to the information displayed in Figures~\ref{monoscale}c and \ref{monoscale}d, a set of Gaussian and coherent transient profiles were numerically generated according to the SNR profile shown in Figure~\ref{multiscale}a, their corresponding \sk random realizations were calculated for several integer multiple of $M=97$, and overlayed on the corresponding \sk profiles (solid lines) and $\esk\pm\sigma_{\ebsk}^2$ ranges (dark grey shaded areas) shown in Figures~\ref{multiscale}c and \ref{multiscale}d, respectively. This comparison demonstrates a very good agreement between the distribution of the \sk random deviates and the theoretical expectations.

As illustrated in Figure~\ref{multiscale}b, although the transients have a fixed duration, their relative duty-cycle increases from $0\%$, up to $\sim91\%$, as an increasing portion of the transient life--time contributes to the accumulation, and it gradually decreases toward $0\%$, as more and more transient free blocks are added to the accumulation. Consequently, the \sk profile of the Gaussian transient (Figure~\ref{multiscale}c) features a double--peak evolution that, for all relative duty-cycle realizations, stays above unity. Although the \sk profile of coherent transient (Figure~\ref{multiscale}d) follows a similar double--peak dependence on the accumulation length, unlike the Gaussian \sk profiles, it reaches less than unity values for the range of accumulation lengths corresponding to relative duty-cycles  above $50\%$.

Therefore, the particular example illustrated in Figure~\ref{multiscale} demonstrates that, if the fixed accumulation length of the \sk spectrometer is tuned to a value that is shorter than twice the expected duration of a coherent transient, while longer than the expected duration of a Gaussian transient, there is a non-zero probability of random realization of an observation that would unambiguously  discriminate the statistical nature of such transients, and thus allow reliable estimations of their SNR and duty-cycles.

Although such particular condition may seem too restrictive for having wide practical applicability, we demonstrate below that it can be straightforwardly achieved by imposing a less restrictive condition on a fixed \sk spectrometer accumulation length, which would be sufficient to be shorter than the duration of both types of transients, in order to provide automatic discrimination capabilities.

Indeed, given the fact that the standard $S_1(M)$ and $S_2(M)$ outputs provided by an \sk spectrometer are additive quantities, they can be sequentially grouped and added together to form the variable length accumulations $S_1(kM)$ and $S_2(kM)$, $k$ being an integer, and thus generate a discreet \sk profile that would follow a continuous accumulation length profile similar to the Gaussian or coherent \sk profiles illustrated in Figures~\ref{multiscale}c and ~\ref{multiscale}d, as it is demonstrated by the numerically generated \sk deviates shown in the same panels.

This concept of \emph{Multi-Scale Spectral Kurtosis} (MSSK) analysis was originally proposed by \citet{sks}, and demonstrated to effectively improve the detection performance of an \sk spectrometer in the case of RFI transients having durations close to half of the fixed accumulation length of the KSRBL instrument. More recently, \citet{jgr} applied the same concept to develop a measurement technique based on fitting the discreet MSSKs profiles with their expected functional forms, which, within the statistical uncertainties, provided estimates consistent with those obtained for the same transient signals based on the mono-scale analysis described in \$\ref{gauss_measurements} and \ref{coherent_measurements}. However, given the fact that, in this particular analysis, the $\sim20$~ms accumulation length the EST instrument was longer than the inferred $\sim8$~ms duration of the observed solar microwave spikes, the corresponding MSSK profiles reproduced only the region around the secondary peak of the expected functional form, and thus the underlaying Gaussian statistics of the solar microwave spikes could not been directly confirmed.

From this perspective, although the classical \sk spectrometer design originally proposed by \citet{rfi} has been proven to already provide, in the form of the $S_1$ and $S_2$ measured quantities, an instrumental output suitable for implementing a downstream real-time MSSK analysis pipeline capable of providing, under favourable circumstances, automatic discrimination of the statistical nature of the observed transients, to generate full length discriminatory MSSK profiles similar to those shown in  Figure~\ref{multiscale}, a modified \sk spectrometer design should be considered.

Such versatile MSSK spectrometer design could involve a hardware implemented continuous computation of \sk estimates that, as the accumulation evolves, would generate Gaussian or coherent \sk flags as soon as one of another transient type is unambiguously identified. If the main goal of such instrument would be the detection and discrimination of transients, the evolving accumulation could be stopped, and the next one initiated, as soon as the transient identification is made, or continued up to a accumulation length that would provide, as demonstrated by Figures~\ref{gauss_estimate} and \ref{coherent_estimate},  transient duration and SNR estimates having the desired level of accuracy.

However, if a fixed accumulation lengthy is preferred, the MSSK spectrometer design could include transient counters that, for each fixed accumulation, would provide two additional data outputs representing counts of detected Gaussian and coherent transients, if any. If the coherent transients are believed to be exclusively generated by RFI, while the coherent transients to be generated by astronomical sources,  this additional information could be subsequently used to safely filter out the accumulations affected by RFI, while preserving the accumulations containing contributions from Gaussian transients shorter than the integration time.
\section{Conclusions}
We obtained analytical expressions that provide biased estimations of the true mean and variance of the \sk distributions in two special cases of spectral transients mixed with a Gaussian time domain background, as functions of their signal-to-noise ratio, and duty-cycle relative o the instrumental accumulation time. We investigated the bias of these approximations and their transient detection performance by means of Monte-Carlo simulations.

We demonstrated that the \sk transient detection performance may be significantly increased, and the bias of \sk--based estimates may be significantly reduced, by increasing the accumulation length. We also developed an analytical workflow leading to estimates of the SNR and duty-cycle of such transients, and their standard-equivalent statistical deviations.

We investigated the accuracy of these estimates and found that, although their statistical uncertainties may vary as function of the SNR and duty-cycle, they can be reduced as low as a few percent by increasing the instrumental accumulation length.

We described a practical adaptive approach that, even for a fixed accumulation length,  may  improve the transient detection performance and reduce the statistical uncertainties of the SNR and duty-cycle estimates, by taking full advantage of the built-in capabilities of the original \sk spectrometer design.

We suggested an original multiscale \sk spectrometer design optimized for real-time detection, classification, and analysis of various transient astronomical signals generated by flaring stars, pulsars, and extragalactic sources, including the elusive Fast Radio Burst transients \citep{FRB}.

Nevertheless, such design may also be considered for the purpose of investigating the existence of coherent natural or artificial astronomical signals, which could be facilitated by a higher order statistics spectrometer as the one proposed here \citep{Melrose}.
\section*{Acknowledgments}
The author tanks the anonymous reviewer for useful comments that helped improve the final version of this manuscript.
\bibliographystyle{mnras}

\bsp
\label{lastpage}
\end{document}